
\magnification=1200
\vsize=22truecm
\hsize=15truecm \tolerance 1000
\parindent=0pt
\baselineskip = 15pt
\lineskip = 1.5pt
\lineskiplimit = 3pt
\font\mybb=msbm10
\def\bb#1{\hbox{\mybb#1}}
\voffset=0pt
\def\la{\langle}
\def\ra{\rangle}

\parskip = 1.5ex plus .5ex minus .1ex
{\nopagenumbers
\rightline{ G\"oteborg-ITP-93-40}
\vglue 1truein
\centerline{A REVIEW OF $W$  STRINGS}
\bigskip
\centerline{Peter West}
\medskip
\centerline{Department of Mathematics}
\centerline{King's College London}
\centerline{Strand, London WC2R 2LS}
\centerline{\&}
\centerline{Institute of Theoretical Physics}
\centerline{Chalmers University of Technology}
\centerline{S-412 96 G\"oteborg SWEDEN}
\bigskip
\centerline{July 1993}
\vskip .75truein
\centerline{Abstract}
\smallskip
Recent progress on the physical states and scattering amplitudes of the $W_3$
string is reviewed with particular emphasis on the relation between this string
theory and the Ising model.
\vfil
\eject}

{\bf 1. Introduction}
\par
I carried out my doctorate studies at Imperial College and had the good fortune
to have Abdus Salam as my supervisor. When I began research, the paper of Wess
and Zumino [33] had induced many of Europe's leading physicists to work on
supersymmetry, and Abdus Salam and John Strathdee [34] had just written their
classic paper discovering superspace and super-Feynman rules. With such rapid
progress being made it was not easy for a graduate student to achieve anything
of significance, but despite his many commitments Abdus Salam was always ready
to give helpful advice and encouragement. It was impossible not to be infected
by his great enthusiasm for new ideas and the enjoyment he derived from doing
physics. One came away from his office feeling that all was possible and that
failure was only a temporary phenomenon.
\par
Although it had been understood [35] how to break supersymmetry using the
classical potential, it was thought to be more desirable if it could be broken
using radiative corrections. Abdus Salam characteristically encouraged Bob
Delbourgo and myself to systematically examine every possibility. He also,
however, advocated that if all else failed one could always tell the truth. In
this case, as I eventually found, the truth was that if supersymmetry was
preserved classically then the effective potential vanished [36]. It was this
theorem which allowed others [37] to observe that supersymmetry solved the
technical hierarchy problem.
\par
The subject of my talk is $W_3$ strings. Paul Howe and myself, and also Bilal
and Gervais, began studying this subject in 1988. The realisation that the
$W_3$ algebra could be used to construct a new string theory was rather
exciting,  since we thought that a much stronger algebra than the Virasoro
algebra must lead to a more exciting string than the bosonic string. After
finding some interesting results, we became despondent. There were two reasons
for this; firstly the only realisation known consisted of two scalars leaving
no room for space-time. Secondly, the construction of the BRST charge by
Thierry-Mieg had an intercept 4 which suggested the possibility of massless
particles of spin greater than 2 in the theory. We realised, however, that
tachyonic
particles with spin necessarily have negative norm states, because we can
choose our frame of reference such that the non-zero components carry a
time-like index. We could not see how to eliminate such particles. No doubt,
had we talked to Abdus Salam we would have continued, but in the event we gave
up.
\par
We now know that $W_3$ strings do not involve massless higher spin
particles and have available representations involving more than 2 scalars. The
message of my talk is that despite their, perhaps, rather artificial appearance
$W_3$ strings show all the magic of ordinary strings, they obey a no ghost
theorem, have scattering amplitudes which obey duality and factorisation and
they are modular invariant. Also one finds that the Ising model pervades all
aspects of the theory.

The discovery, by Zamolodchikov [1], of two dimensional algebras that contain
currents of spins greater than two has led to many interesting new
developments.
The simplest such algebra, $W_3$  is generated by the spin 2 energy-momentum
tensor $T(z)=\sum L_n z^{-n-2}$ and the  spin 3 current $W(z)=\sum W_n
z^{-n-3}$; the algebra being
$$
\eqalign{
[L_n, L_m]&=(n-m)L_{n+m}+{c\over 12}n(n^2-1)\delta_{n+m,0}\ ,\cr
[L_n, W_m]&=(2n-m)W_{n+m}\ ,\cr
[W_n, W_m]&={16\over{22+5c}}(n-m)\Lambda_{n+m}\cr
&\ \ +(n-m)\Big[{1 \over{15}}(n+m+2)
(n+m+3)-{1\over 6}(n+2)(m+2)\Big]L_{m+n}\cr
&\ \ +{c\over 360}n(n^2-1)(n^2-4)\delta_{n+m,0}\ ,\cr}
\eqno(1.1)
$$
where
$$
\Lambda_n=\sum_m :L_{n+m}L_{-m}:-{1 \over20}\Big[n^2-4-{5 \over 2}\big(1-(-1)^n
\big)\Big] L_n
\eqno(1.2)
$$
and the normal ordering means that $:L_pL_q:=L_qL_p$ if $p > q$.
\par
This algebra occurs naturally in the minimal model with central charge 4/5
since this theory contains a spin 3  primary field. The above algebra, however,
is consistent for all values the central charge $c$. The most important new
feature of these new algebras is that they are not Lie algebras; their
commutators are expressed as a sum of terms some of which are quadratic in the
generators.
\par
 An important consequence of this fact is that, unlike the Virasoro
algebra, two representations of a $W_3$ algebra cannot, in general, be added to
form a third. The earliest known realisation of the $W_3$ algebra is in terms
of 2 scalars [2]; it is a generalisation of the classical Miura transformation
that takes the mKdV equation into the KdV equation. It was found [3], however
that one of the scalars only occurred through its energy-momentum tensor and
consequently, one could replace it by any energy momentum tensor with the same
central charge. In particular, one could use any number of scalar fields, say $
x^\mu$ for  $\mu = 0,1,..., D-1$ with a suitable background charge. Denoting
the other of the two original scalars by $\varphi$, the $W_3$ generators, which
carry a superscript m, corresponding to matter, are given by
$$
T^m  = T^\varphi + T^x \eqno(1.3)
$$
where
$$
\eqalignno{
T^\varphi & = -{1\over 2}{(\partial\varphi)}^2 -
  Q \partial^2\varphi,&(1.4)\cr
T^x & = -{1\over 2} \partial x^{\mu} \partial x_{\mu}
  - \alpha_{\mu}\partial^2 x^{\mu}, & (1.5)\cr}
$$
and
$$
W^m  =- {2i\over \sqrt{261}}\left[{1 \over 3} { (\partial \varphi)}^3 +
  Q \partial\varphi \partial^2 \varphi
  +{1 \over 3} Q^2 \partial^3 \varphi +
  2 \partial\varphi T^x + Q \partial T^x\right], (1.6)$$
\par
It was natural given the existence of the $W_3$ algebra to try to build new
string theories.   Although
there are,  with hindsight,  a number  of ways of constructing
the bosonic  string, the  path  most  suited to our present knowledge of the
$W_3$ algebra is as follows: starting from the Virasoro algebra we
construct the  BRST charge $Q$ and demand  its square vanish.
This condition implies that c = 26 and the intercept is one [4].  We
then find  a realization  of the Virasoro algebra with c = 26,
for example  26 free  scalars $x^{\mu}    ,  \mu =  0, 1,...,  25$.   The
physical states are the non-trivial cohomology classes of Q  which are
also subject  to a ghost constraint [4,5,15]. Given the physical
states, one can then construct the
scattering amplitudes.
\par
The construction of a $W_3$  string along
these lines  was first discussed in reference [6] and the $W_3$ string with the
field content given here was first formulated in reference [7].  Fortunately,
the  BRST charge for the $W_3$
algebra  had previously  been constructed [8]
and found to square to zero if c = 100 and the intercept was 4. To build this
operator one must introduce the usual reparameterization
ghosts $b$ and $c$, and $W_3$ transformation ghosts $d$ and $e$ which have
spins $3$ and $-2$ respectively. The total energy-momentum tensor $T^{tot}$ and
spin 3 current $W^{tot}$ of the combined ghost and matter system are
of the form
$$
T^{tot} = T^m + T^{gh} , \quad W^{tot} = W^m + W^{gh} \eqno(1.7)
$$
where
$$\eqalignno{
T^{gh} &  = - 2 b\,\partial c - \partial b\,c - 3d\, \partial e -
2 \partial d\, e , &(1.8)\cr
W^{gh} & = - \partial d\, c - 3 d\, \partial c -{8\over 9.27} [\partial
(b\,e\, T^m) + b\,\partial e\, T^m]\cr
& + {25\over 6.9.27} (2 e\, \partial^3 b + 9 \partial e\, \partial^2 b
  + 15 \partial^2 e\, \partial b + 10 \partial ^3 e\, b). &(1.9)
}
$$
Here the background charge $Q$ is given by $Q^2 = 49/8$, and $\alpha$ is
such that $T^x$ has central charge $51/2$.
The BRST charge $Q$ is given by [8]
$$
Q = \int dz \, j^{BRST}, \eqno (1.10)
$$
where
$$
j^{BRST} = c(T^m + {1\over2} T^{gh}) + e(W^m + {1\over 2} W^{gh}).\eqno (1.11)
$$
The reader will be aware of the difference between the background charge $Q$
and the BRST charge $Q$ from the different contexts in which they are used.
Some useful relations are
$$
T^{tot}(z) = \left\{ Q, b(z)\right\} , \quad W^{tot} (z) =
\left\{Q, d(z) \right\}, \eqno (1.12)
$$
as a consequence of which
$$
[Q, T^{tot}(z)] =  \left [Q, W^{tot}(z)\right ] = 0. \eqno (1.13)
$$
\par
It will be useful to discuss the various possible vacua associated
with the ghosts.  The natural vacuum, $\vert 0\rangle$, of the ghost system is
that for which $e(z) = \sum_n e_{-n} z^{n+2}$ and
$d(z) = \sum \limits _n d_{-n} z^{n-3}$ are well defined at $z = 0$.
This requires [9]
$$
\eqalign{
e_n \vert 0\rangle &= 0, \quad n\ge3,\quad  d_n \vert 0\rangle = 0,\quad n\ge
-2\cr
c_n \vert 0\rangle &= 0, \quad n\ge 2,\quad b_n \vert 0\rangle = 0, \quad n\ge
-1.\cr}
\eqno(1.14)
$$
We can construct other vacua by acting on $\vert 0\rangle$ with $e_n$
for $n=0,\pm 1, \pm 2$ and with $c_n$ for  $n=0, \pm 1$. One
of the most useful is
$$
c_1 e_1 e_2 \vert 0 \rangle \equiv  \vert \downarrow \rangle, \eqno (1.15)
$$
which is annihilated by $e_n, c_n$ for $n \ge 1$ and $b_n, d_n$ for
$n\ge 0$.  In
terms of the conformal fields we may express the relation between
the two states as
$$
c\,\partial e\, e \vert 0 \rangle = \vert \downarrow \rangle, \eqno (1.16)
$$
where $c\, \partial e\, e$ is understood to be evaluated at $z =0$.
Similar formulae hold for the other vacuum states.
\par
In order to gain a non-zero vacuum expectation value with respect to the
state $\vert 0\rangle$ we must have 3 factors of $c$ and 5 of $e$. We set
$$
\la 0\vert c_{-1}c_0 c_1 e_{-2} e_{-1} e_0 e_1 e_2 \vert 0\rangle  =
{1\over 576} \la 0\vert \partial^2 c\, \partial c\, c
\partial^4 e\, \partial^3 e\,
\partial ^2 e\, \partial e\, e\vert 0\rangle  = 1 \eqno (1.17)
$$
A correlator will also vanish unless the sum of the momenta of all the vertex
operators in the correlator is $2i$ times background charge.

  {\bf   2. States in effective bosonic strings}

In $W_3$ and probably $W_N \ N \geq 4 $ strings one finds that the physical
states belong to only  a subspace of
the full Fock space of the theory that is similar to the bosonic string. In
this section, we will consider what are the physical states that can arise in
such effective bosonic string sectors, but rather than base our discussion
specifically within the context of $W$ strings we will consider bosonic
string-like sectors in the abstract, so making the discussion clearly
applicable to any string theory in which they may arise. By an effective
bosonic string sector we mean a string, or sector of a string theory, that has
$D$ scalar fields $x^{\mu}$ which possess
a background charge $\alpha ^{\mu} $. The physical states of the sector belong
to the Fock space subspace $\tilde H$ generated by $\alpha^{\mu}_n$ and are
taken to be subject to the Virasoro-like conditions
$$ \tilde L_n|\tilde \psi \rangle =0 , \ n  \geq 1 \ \ ,
(\tilde L_0 - a)|\tilde \psi \rangle =0 \eqno(2.1)$$
where the intercept $a$ is to be specified and
$$ \tilde T = -{1 \over 2} \partial x^{\mu} \partial x^{\nu} \eta_{\mu \nu}
 - \alpha_{\mu}{ \partial}^2 x^{\mu} \eqno(2.2)$$
$$\tilde T(z)=\sum_n \tilde L_n z^{-n-2},\ \  i\partial x^{\mu}(z) =\sum_n
\alpha^{\mu}_n z^{-n-1} \eqno(2.3)$$
We note that the background charge need not, be such that $c=D+12{
\alpha}^2 $ is equal to the critical value of 26.
\par
The physical states in such a theory were found in reference [10] by using an
extension of the methods of reference [11] and we now summarise these
arguments. The first step is to introduce a set of operators that span $\tilde
H$, but have sufficiently simple commutation relation properties with $\tilde
L_n$ as to allow us to  solve the constraints of equation (2.1). We begin by
defining the operators
$${\cal A}^{\mu}_n = A^{\mu}_n -{{n k^{\mu}-2i\alpha^{\mu}} \over 2}  F_n
\eqno(2.4) $$
where
$$ A^{\mu}_n =\oint dz :i\partial x^{\mu} e^{ink\cdot x }:  $$
$$ F_n = \oint dz :{k \cdot \partial \partial x \over k \cdot \partial x}
e^{ink\cdot x} : \eqno(2.5) $$
These will commute with $\tilde L_n$  provided $k \cdot (k- 2i\alpha) =0$. If
we
further take $k^2=0$,    then they obey the algebra.
$$[{\cal A}^{\mu}_n ,{\cal A}^{\nu}_m ] =\left( n\eta ^{\mu \nu }+2n^3k^\mu
k^\nu \right) p.k\delta _{n+m,0}+mk^\mu {\cal A}^\nu _{n+m}-nk^\nu {\cal A}^\mu
_{n+m}$$
$$ -i\alpha ^{\mu} k^{\nu} m^2 \delta _{n+m,0} p\cdot k +i\alpha ^{\nu} k^{\mu}
m^2 \delta _{n+m,0} p\cdot k . \eqno(2.6)$$
\par
We next define the operators
$$ C_n = - {\cal A }_n^{-} -\left\{ {1 \over 2} \sum _p \sum _i:C^i_{n-p}C^i_p:
-i(n+1) \alpha \cdot C_n \right\} +1$$
$$C_n^i = A_n^i +i \alpha ^i \delta _{n,0} \eqno(2.7) $$
which of course also commute with $\tilde L_n$.
These operators obey the algebra
$$[C^i_n,C^j_m]=\delta _{n+m,0}\delta _{i,j} ,\ [C_n^i ,C_m ] =0
,$$
$$ [C_n,C_m] =(n-m)C_{n+m} +(26-D-12 \alpha ^2){n(n^2-1) \over 12
}\delta _{n+m,0} \eqno(2.8) $$
\par
Since $C_n$ and $C_n^i$ commute with $\tilde L_n$, we
conclude  that they  generate  physical states which are of the form
$$C_{-m_1}.....C_{-m_q} C^{i_1}_{-n_1}....C^{i_p}_{-n_p} |0,p \rangle
\eqno(2.9)$$
where $ |0,p \rangle$ is  the  tachyon  state  and is annihilated  by
$\alpha_n^{\mu},\ n \ge 1$ and $p^{\mu}$ satisfies $p \cdot k=1,\ {1 \over 2} p
\cdot (p-2i\alpha )= a$.
\par
We add to
our collection of oscillators the
$$\phi _n=\oint {dz} z^{-1}{\rm e}^{ink.x}\ \ . \eqno(2.10) $$
These have the relations
$$[\tilde L_p,\phi _m]=-p\phi _m\ \ ,\ \ [C_p,\phi _m]=m\phi _{m+p}$$
$$[\phi _n,\phi _m]=0\ \ ,\ \ [C^i_n,\phi_p]=0\ . \eqno(2.11) $$
One can  show that  the oscillators $C^i_n$,  $C_n$ and $\phi _n$ do span the
same Hilbert  space as  the original  oscillators $\alpha ^\mu _n\ ,\ \mu
=0,1,\dots ,D-1$. It also follows, from the above commutation relations, that
the physical states do not contain $\phi _n$ and so are none other than those
of equation (2.9)

We now wish to analyse the physical states. The operators $C_{-n}$ obey a
Virasoro algebra which has a
central charge $\tilde c=26-D-12 \alpha ^2$ and they act on highest weight
states with
weight $h =1-a$. The norms of these states is therefore controlled by the Kac
determinant for the Virasoro-like operators $C_n$ with the above central charge
and weight. If $\tilde c $ is less than or equal to 1 then the states will only
have positive norm  if and only if $\tilde c $ is a member of the minimal
unitary series
 $\tilde c =1-{6 \over {(n+1)(n+2)}}\ n=3,4,...$ and the intercepts $a_{r,s}$
are related to the weights of the corresponding minimal unitary series by
$a_{r,s}=1-h_{r,s}$.
The number of physical degrees of freedom is found by discarding those states
of zero norm. This is the same as discarding those states which are descendant,
but yet also highest weight with respect to the $C_n$.  Consequently, the
number of physical degrees of freedom $c_n$ at
level  $n$  is given by
$$ \sum c_n x^n = \prod _{n=1}^{\infty} {1 \over (1-x^n)^{D-2}} \hat \chi
_{h_i}(x) .
\eqno(2.12) $$
The character $\chi _h $ is that corresponding to the above central charge and
weight and is defined by
$$\chi _h(z)=\sum_{n=0}^\infty z^{h-{1\over 48}}{\rm dim}V_{n+h}=z^{h-{1\over
48}}\hat \chi _h (z) \eqno(2.13) $$
where $h$ is the weight of the highest weight state and $V_q$ is the dimension
of
the space with weight $q$.
\par
For a critical bosonic string  $ c =26 $ and $a=1$, or equivalently  $\tilde c
=0 $ and $h=0$, and in this case all the states involving $C_{-n}$ are null and
so one finds that the count of states is a light-cone count in the sense that
there are excitations for only $D-2$ of the $D$ non-zero modes.  When the
effective bosonic string, or sector, is not critical one finds that the string
does not  lose 2 degrees of freedom and
only some of the states involving $C_{-n}$ are null. When one is dealing with
one of the minimal models, which must be the case if $\tilde c \leq 1$ and the
states obey a no ghost theorem, the explicit form of the characters is known
[12] and so equation (2.12) gives the count of states. In this case, one can
then consider the question of whether the string theory is modular invariant.
The discussion of reference [10] is for the Ising model case, but it is trivial
to extend it to the general case. One finds that the cosmological constant is
the product of a factor that is associated with the $x^{\mu}$ oscillators,
which is modular invariant by itself and a factor associated with the minimal
model. For the theory to be modular invariant this latter factor must also be
invariant. Consequently, for every modular invariant of the minimal series [13]
one finds a corresponding modular invariant string.

{\bf 3. The physical states of the $W_3$ string}

Before we begin, it will be useful to recall what are the physical states in
the    ordinary critical bosonic string constructed from the 26 fields
$x^{\mu}$. These states were found, by studying the original dual model [14],
to obey the Virasoro constraints $(L_0-1)\vert\psi^{x} \rangle=0$,
$L_m\vert\psi^{x} \rangle=0 \ m \geq 1$ where $\vert\psi^x\rangle$ depends only
on $x^\mu$. With the discovery of the free gauge covariant action, it became
more common to consider the physical states as belonging to the cohomology of
the BRST operator $Q$, subject to a certain ghost number constraint.  We recall
that the cohomology of $Q$ [4,15]
consists of the states $\vert\psi^x\rangle\vert\downarrow \rangle$ and
$\vert\psi^x\rangle \,c_0\vert\downarrow \rangle$,
where $\vert\downarrow\rangle=c_1\vert 0\rangle$ and $\vert 0\rangle$ is
the $SL(2,\bf{R})$ invariant vacuum, as well as   two further states
with zero momentum, namely $\vert p^{\mu}=0 \rangle \vert 0\rangle$ and $\vert
p^{\mu}=0 \rangle c_{-1}c_0c_1\vert
0\rangle$.

For a $W_3$ string, it is also natural to  regard the physical states as being
given by the cohomology of $Q$, subject to a suitable ghost number
constraint. By analogy, we expect physical states to be contained in states of
the form $\vert\psi^{x,\varphi}\rangle\vert\downarrow\rangle$, where
$\vert\downarrow \rangle=c_1e_1e_2\vert 0\rangle$. Applying $Q$ to these states
one finds [6] that they belong to the cohomology of $Q$ if
 $$
\eqalign{(L^m_0-4)\vert\psi^{x,\varphi} \rangle=&0,\quad
W^m_0\vert\psi^{x,\varphi} \rangle=0,\cr
L^m_n\vert\psi^{x,\varphi} \rangle=&0,\quad
W^m_n\vert\psi^{x,\varphi}\rangle=0,\quad n\geq 1.\cr}\eqno(3.1)
$$

Included amongst the states satisfying these conditions
are physical states having the form [7]
$$
\vert\psi^{x,\varphi}\rangle = \vert\psi^x\rangle
\vert 0,\beta\rangle\vert\downarrow\rangle,\eqno(3.2)
$$
where $\vert 0,\beta \rangle$ is a state with $\varphi$ momentum
equal to $\beta$ and
no $\varphi$ oscillators.  Such states will satisfy the conditions (3.1)
provided that the state $\vert\psi^x\rangle$, which
depends on $x^\mu$ alone, satisfies the conditions
$$
L^x_n\vert\psi^x\rangle=0,\ n\geq 1,\quad(L^x_0-a)\vert\psi^x\rangle
=0,\eqno(3.3)
$$
where $a=1$ for $\beta={8iQ/7}$ and ${6iQ/7}$, and
$a={15/16}$ for $\beta=iQ$.  The above values of $\beta$ are
in fact the only ones allowed by the on-shell conditions of equation
(3.1). These states exhibit what is a general phenomenon, namely the freezing
of the $\phi$ momentum by the physical state conditions which was first noticed
in reference [7]. The reader may observe that if one takes one minus the above
effective intercepts then one gets 0 and 1/16 which remind one of the weights
that occur in the Ising model. In references [7] and [16] this, and other
phenomenological number matching, was shown to extend to a relation between
higher $W_N$ strings and minimal models.
\par
Further analysis of low level states was carried out in references [17]and
[18], but it was only in reference [19] that a systematic study of the physical
states of the type of
equation (3.1) at levels up to and including 2 was undertaken.  It was found
that any state that contained non-zero mode
$\varphi$ oscillators acting on the vacuum was null. By examining all other
null states, the count of physical degrees of freedom at these
levels was found.  It thus became clear that the non-null physical
states based on the ghost state $|\downarrow \rangle $ were of the form of
equation (3.2) and that  the open $W_3$
string had only one massless particle, a photon, namely the same massless
states as the open bosonic string [19].
\par
It will be useful to rewrite the physical states discussed above in terms of
vertex operators acting on the vacuum $|0 \rangle$ in the ghost sector as well
as the usual vacua for the bosonic oscillators. The states of equation(3.2)
correspond to the vertex operators
$$
\eqalignno{V(1,0)& = c\,\partial e\, e\
e^{i\beta(1;0)\varphi}V^x(1)&(3.4)\cr
\bar V(1,0)& =\ c\,\partial e\, e\
e^{i\bar\beta(1;0)\varphi}V^x(1),&(3.5)\cr}
$$
where, with further developments in mind, we have introduced the notation
 $\beta (1;n) = (8-8n) {iQ\over7}$ and $\bar \beta (1,n) = (6-8n) {iQ\over 7},
$
and
$$
V({15/16},0) =\ c\partial e\ e\
e^{i\beta({15/16},0)\varphi}V^x({15/16})\eqno(3.6)
$$
where $\beta ({15/16}, n) = ({7-4n}) iQ/7$.
Here $V^x(a)$ is any vertex operator, constructed from $x^\mu$
alone, that has conformal weight $a$ with respect to $T^x(z)$.  The
simplest is $V^x(a)=e^{ip\cdot x}$, where ${1\over 2}p\cdot(p-2i\alpha)=a$.
\par
For the bosonic string, all the physical states, except for those of fixed
momentum, are built on a single state with given ghost number, but this is
not the case for the $W_3$ string. States in the cohomology of $Q$ which are
not of the above form were first found in the context of the 2 scalar $W_3$
string in reference [16] and one such state was found in the 3 scalar $W_3$
string. It was shown, as discussed below, that the consistency of the $W_3$
scattering [21] and modular invariance [10] also required
further states to those listed above and these additional states are given by
[10]
$$
V({1/2},0) = \biggl(c\,e-{i\over \sqrt{522}} \partial
e\,e\biggl)\ e^{i\beta({1/2},0)\varphi}V^x({1/2}) \eqno(3.7)
$$
where $\beta ({1/2}, m) = (4-8m){iQ\over 7}$.

These states were found by a vanishing null state argument which also led to
states in the cohomology of $Q$ that were of the form [10]
$$
V({15/16},1) =\ \biggl(c\,e+{i\partial
e\,e\over\sqrt{522}}\biggr)\ e^{i\beta({15/16},1)\varphi}
V^x({15/16}).\eqno(3.8)
$$
and
$$
\bar V({1/2},0) = \biggr(
-{4\over 3 \sqrt{58}}b\,c\,\partial e\,e
  -{4\over 3 \sqrt{58}}\partial^2 e\,e
  + {1\over\sqrt{29}} \partial\varphi\partial e\,e
  + i \sqrt 2 c\,e\,\partial\varphi
  -{3i\over2} c \partial e\biggl)
$$
$$
\times e^{i\bar\beta({1/2},0)\varphi}V^x({1/2}),\eqno(3.9)
$$
where
$\bar \beta ({1/2}, m) = (2-8m){iQ\over 7}$.
This argument also implied the presence of an infinite number of other such
states.
The above physical states of the $W_3$ string of equations (3.4), (3.6) and
(3.7) are in effect contained in sectors which are effective bosonic strings
with intercepts 1,15/16 and 1/2 and central charge $ c=25+1/2$. The states of
equations (3.5), (3.8)and (3.9) are also of this form with effective intercepts
1, 15/16 and 1/2 respectively and so are in effect only copies of the previous
states.  Further such copies were found in references [22,26]
\par
We can now analyse the physical content of these states, using the results of
the previous section. These physical states are of the form of equation (2.9)
with $\tilde c=1/2$ and highest weights $h$ of 0,1/16 and 1/2 and so the count
of states of equation (2.12) has characters which are none other than those of
the Ising model and further, these states obey a no ghost theorem in that they
have positive norm [10].
We note that given $\tilde c ={1 \over 2}$ these are the only values of the
intercepts for which this is true. Thus, although the construction of the $W_3$
string did not seem to involve the Ising model, we find a deep connection
between the states of the two theories.
\par
We can now consider whether the above states are sufficient in number for the
cosmological constant to be modular invariant. Since the partition function
involves the Ising model characters one might suspect that one finds a modular
invariant $W_3$ string for every modular invariant  Ising model. In fact there
is only one such Ising model and so only one modular invariant $W_3$ string. We
refer the reader to reference [10] for a detailed discussion.

These facts, and the result, discussed below, that the factorisation of the
tree level $W_3$ scattering amplitudes for these states leads to no new states
strongly suggests [10] that the cohomology of $Q$ involves the states of
equations (3.4),(3.6) and (3.7),copies of them, and possible discrete states.
\par
Considerable further evidence for this conjecture was provided in reference
[23] where it was shown how to generate infinite classes of states belonging to
the cohomology of $Q$. We now summarize this work. The existence of copies
suggests that there should be a screening charge of the form
$$\int dw\ e^{i\beta\varphi}f(b,c,d,e,\partial\varphi)\eqno(3.10)$$
that commute with $Q$ and takes one copy into another. If we can find such
charges that do not involve the field $x^\mu$,
then the insertion of these charges into a correlation function will not
change the effective space-time interpretation of the correlation
function. Indeed three such screening charges do exist [23] and are given by
$$\int dz\ e\ e^{i\beta\varphi} \eqno(3.11) $$
for $\beta={6iQ/7}$ or $\beta={8iQ/7}$.
and, more importantly for our present purposes, the operator
$$
S = \oint dz \{d-{5i\over 3\sqrt{58}} \partial b-{2\over
3.87} \partial b\,b\,e-{4i\over 3}{1\over \sqrt{58}} d\,b\,e
\}e^{i\beta^s\varphi}\eqno(3.12)
$$
where $\beta^s=-{2iQ/7}$.
Since $S$ commutes with $Q$, it follows that whenever the action of
$S$ on a physical state is well-defined and non-zero it will
produce another physical state.
\par
Let us now consider when two screening operators have a well
defined action on a vertex operator with $\varphi$ momentum
$\beta$.  Since this has not, to our knowledge, been clearly
discussed in the literature let us consider the more general case
of two screening operators with $\varphi$ momentum $\beta^s_i,
i=1,2$.  Since the factors in front of the exponentials are made
from either ghosts or $\partial\varphi$ they can only contribute
integer powers of the coordinates in the operator product expansion, and
consequently do not affect
whether or not the action of the screening operators is well
defined.  Consequently, we must focus our attention on the factors
$$
\oint\limits_{C_1}dw_1\ \oint\limits_{C_2}dw_2\
 e^{i\beta^s_1\varphi(w_1)}\ e^{i\beta^s_2\varphi(w_2)}\
  e^{\beta\varphi(z)}
$$
$$ = \oint\limits_{C_1}dw_1\
  \oint\limits_{C_2}dw_2\ (w_1-w_2)^{\beta^s_1\beta^s_2}(w_1-
  z)^{\beta^s_1\beta}(w_2-z)^{\beta^s_2\beta} \
e^{i\beta^s_1\varphi(w_1)+i\beta^s_2\varphi(w_2)+i\beta\varphi(z)}
\eqno(3.13)
$$
Since $\vert w_1-z\vert > \vert w_2-z\vert $ we arrange the
$w_1$ and $w_2$ contours to be around the point $z$ in such a way
that the above condition is satisfied.

Let us consider the substitution $(w_1,w_2)$ to $(w_1,\tau)$ given
by $(w_2-z)=\tau(w_1-z)$.  The above condition implies that
$\vert\tau\vert< 1$, but we also demand that $\tau=1$ at one and
only one point in other words the $w_1$ and $w_2$ contours touch one
another once.  This latter condition ensures that the value of the
integral is dependent on only the one place where the contours
cross the branch cut.  We are to regard $w_2$ as fixed and consider
the $\tau$ integration.  Substituting $dw_2=d\tau(w_1-z)$ we find
the above integrals become
$$\eqalign{\oint_z dw_1\oint d\tau
&\tau^{\beta^s_1\beta}(1-\tau)^{\beta^s_1\beta^s_2}(w_1-z)^P\cr
&\exp(i\beta^s_1\varphi(z+(w_1-z))+
i\beta^s_2\varphi(z+\tau(w_1-z))+i\beta\varphi(z))\cr}\eqno(3.14)
$$
where $P=1+\beta^s_1\beta+\beta^s_1\beta^s_2+\beta^s_2\beta$.  This
integral is well defined if $P$ is an integer.

The generalization
to $n$ screening charges with $\varphi$ momenta $\beta^s_i$ is
straightforward.  Their action contains the term
$$
\eqalign{
\prod\limits^n_{i=1}\left(\oint dw_i\,
  e^{i\beta^s_i(w_i)}\right) e^{i\beta\varphi(z)} = & \cr
\oint
  dw_i\prod\limits^n_{{i<j}\atop {i,j=1}} (w_i-
  w_j)^{\beta^s_i\beta^s_j}\prod\limits^n_{j=1} (w_j-
  z)^{\beta^s_j\beta}
& \exp i\{\sum\limits^n_{j=1}\beta^s_j\varphi(w_j)+\beta\varphi(z)
  \}\cr
}\eqno(3.15)
$$
We now replace $w_j$, $j=2,\dots,n$ by $w_1,\tau_j\ j=2,...,n$
using the formula $(w_j-z)=\tau_j(w_1-z)$.  The above expression
becomes
$$
\oint\limits_z\ dw_1\prod\limits^{n-1}_{i=1}\oint d\tau_i\
f(\tau_i)\ (w_1-z)^{P^\prime}\ \exp
i\left\{\sum\limits^n_{j=1}\beta^s_j\varphi(z+\tau_j(w_n-
z))+\beta\varphi(z)\right\}\eqno(3.16)
$$
where $f(\tau_i)$ is a function of $\tau_i$ and
$$
P^\prime\ =\ (n-1)+\sum\limits^N_{{i,j=1}\atop{i<j}}\
\beta^s_i\beta^s_j + \sum_{j=1}^N \beta^s_j\beta.\eqno(3.17)
$$
Thus the integrals are well defined if $P^\prime$ is an integer.

Let us apply this general discussion to the case of interest,
namely the action of $n$ screening charges $S$ with momentum
$\beta^s=-{2iQ/7}$.  Taking $\beta={isQ/7}$ with $s$ an integer, their
action is well defined if
$$
{n\over 4}[-(n-1)+s] \in \bb{Z},\eqno(3.18)
$$
and then the momentum of the resulting vertex is
$$
{iQ\over 7}(s-2n).\eqno(3.19)
$$
For the action of screening charges on $V({15/16},0)$ we have $s=7$. We then
find that $n$
must be even and that the vertices so constructed have momenta
$
\beta({15/16},m)\ =\ {iQ\over 7}(7-4m),\quad
m \in \bb{Z}_+ .
$
For the action on $V(1,0)$, for which $s=8$, we find that $n=4m$ or
$4m+1$ with $m \in \bb{Z}_+$, which leads to the vertices with
momenta
$
\beta(1,m) = (8-8m){iQ\over 7},\quad
m \in \bb{Z}_+,$
and
$
\bar\beta(1,m) = (6-8m){iQ\over 7}\quad,
m \in \bb{Z}_+.$

Finally, the action of $n$
screening charges on
$V({1/2},0)$ is well defined if $n=4m$ or $4m+1$ with
$m \in \bb{Z}_+$, which leads to vertices with momenta
$
\beta({1/2},m)\ =\ {iQ\over 7}(4-8m)$
and
$
\bar\beta({1/2},m)\ =\ {iQ\over 7}(2-8m).$

It can, and does happen, that applying $S$'s alone leads to a
vanishing result.  This can be avoided, however, by the judicious
use of the picture changing operator $P$ which is of the form $P=[a\cdot
x+\varphi,Q]$; this operator was first used in the context of $W_3$ strings in
reference [22].

The general pattern is as follows; in the intercept $15/16$ sector we have the
vertices
$$
V({15/16},m)\ =\ (S^2P)^m\ V({15/16},0)\eqno(3.20)
$$

while for the intercept $a=1$ and 1/2 vertices, we have the vertices
$$
V(a,m)\quad {\rm and}\quad \bar V(a,m)\eqno(3.21)
$$
These latter vertices are defined by the relations
$$
\bar V(a,m)\ =\ SPV(a,m),\eqno(3.22)
$$
$$
V(a,m)\ =\ S^3P\bar V(a,m-1).\eqno(3.23)
$$
\par
To summarise, we have found [23] that given the basic vertices
$V(a,0)$ for $a=1$, ${1/2}$ and ${15/16}$, we can use $S$ and
$P$ to create the BRST invariant vertices $V(a,m)$; for $a=1,1/2$
we obtain in addition the vertices $\bar V(a,m)$.
We can also obtain further vertices by the action of $P$
on these. Since the physical states for a
given intercept have a spectrum generating algebra involving the
operators $C^i_n,i=1,\dots D-2$ and $C_n$ the states in the cohomology of $Q$
are generated by
$C^i_n,C_n,S$ and $P$.

A number of examples of this procedure were given in reference [23], but here,
we give only one example relating the first two copies in the intercept 1/2
sector.  A short calculation shows that
although $SV({1/2},0)$ is well defined it vanishes.  We
therefore introduce further powers of the ghost $e$ into the
vertex using the picture changing operator. We then have
$$
\eqalign{
PV({1/2},0)= & \bigl( 5 \,\partial^2 e\,e -
  {24 Q\over 7} \partial\varphi \partial e\,e\cr
&  - {19 i\over 3\sqrt{58}} \partial^2 e\,\partial e\,e
\bigr)e^{i\beta(1/2,0)\varphi}V^x(1/2),\cr
}\eqno(3.24)
$$
and acting on this with the screening operator we find the
vertex $\bar V({1/2},0)$,
$$
SPV({1/2},0)\propto\bar V({1/2},0).\eqno(3.25)
$$

{\bf  4. $W_3$ String Scattering}
\par
The scattering, at tree
level, of $W_3$ strings was first found using the group theoretic method [24].
It was found that these scattering amplitudes contained within them the Ising
model correlation functions, in particular the scattering amplitude of $N$
$W_3$ string states is given by [21]
$$
\int {\prod \limits_ i}^\prime dz_i \,V f(z_i).\eqno(4.1)
$$
Here $V$ is the usual scattering vertex in the presence of a
background charge, and $f$, the measure, is an Ising model correlation function
that depends on the intercepts of the external states.  To be
specific, if the $N$ external states have effective intercepts
$a_i,\quad i=1, ...,N$, which can take only the values $1$,
${15/ 16}$ or ${1/ 2}$, then $ f = \la \prod \limits^N _{i=1}
\varphi_i (z_i)\ra$ where $\varphi_i$ is the
Ising field of weight $h_i = 1 - a_i$.
\par
The essential steps, using the group theoretic method to  calculate a
scattering amplitude are first the computation of the vertex
$V$ using overlap relations, and then the determination of the
measure $f$ by demanding that null states decouple. Using this
technique it was possible to work with the reduced subspace of the
full $W_3$ Fock space in which the physical states sit.  It is
important to understand that the properties of the vertices and
null states used in this reduced Hilbert space are those inherited
from the full Fock space of the $W_3$ string.
We found that the decoupling of the null states of the $W_3$ string
implied that $f$ obeyed the differential equations satisfied by the
Ising model. The reader is referred to reference [21] for the details of this
derivation.

It is straightforward to evaluate the $W_3$ string scattering
given by equation (3.1) whenever the Ising model correlation
functions are known. The vertex $V$ can be found in reference [21] for any
states of the $W_3$ string, but for tachyons it has the simple expression
$$
\prod \limits _{i<j} ( z^i - z^j )^{2\alpha^\prime p_i \cdot p_j}
\eqno (4.2)
$$
For four tachyon scattering, with the choice of Koba-Nielsen
coordinates $z_1 = \infty$, $z_2 =1$, $z_3 = x$ and $z_4 = 0$,
this reduces to
$$
x^{-\alpha^\prime s - a_3 - a_4} (1 - x)^{-\alpha^\prime
t - a_2 -a_3}, \eqno (4.3)
$$
where $s = -(p_1 + p_2)\cdot(p_1 + p_2 - 2{i\alpha})$,
$t = -(p_2 + p_3) \cdot(p_2 + p_3 - 2{i\alpha})$ and where a
factor of $(z_1)^{-2a_1}$,
which is cancelled by other such factors in the amplitude, has been removed.
\par
Let us give one example of the evaluation of such amplitudes, namely the case
of four intercept 15/16 tachyonic $W_3$ strings, whose subsequent study will be
instructive.
$$
F(s,t) = \int_0^1 d x x^{-\alpha' s - 15/8} (1-x)^{-\alpha' t-15/8} f(x)
z_1^{1/8} \eqno(4.4).
$$
The function $f(x)$ is the Ising correlation function for 4 weight 1/16 fields
which
is of the   form [25]
$$
f(x) = [ (z_1 - z_3)(z_2 - z_4)]^{-1/8} Y(x) \eqno(4.5),
$$
where
$$
x={(z_1-z_2)(z_3-z_4)\over (z_1-z_3)(z_2-z_4)}
 \eqno(4.6)$$
and
$$
Y(x) = {1\over[x(1-x)]^{1/8}} (a \cos\theta + b \sin\theta),
\eqno(4.7)$$
with $x=\sin^2\theta$.  The two constants $a$ and $b$ correspond to the two
solutions to the second order differential equation that this correlation
function satisfies due to the existence of null states in the Ising model.
Taking the above choice of $z$'s
we find that
$$
F(s,t) = \int_0^1 d x x^{-\alpha's-2} (1-x)^{-\alpha' t-2}(a \cos\theta + b
\sin\theta). \eqno(4.8)
$$
In the above we have introduced
the slope
$\alpha'$,
which is usually taken to be $1/2$ for the open string.

The values of the constants $a$ and $b$ are to be chosen by the physical
requirement of crossing. The open string amplitude $T^{(4)}(p_i)$ is as usual
the
sum of three
terms
$$
T^{(4)}(p_i) = F(s,t) + F(t,u) + F(u,s).
 \eqno(4.8)$$
Crossing for four identical particles means that $T^{(4)}(p_i)$ should be
symmetric under the
exchange of any two legs or equivalently momenta.  This means, for example,
that
it should be
symmetric under $s\leftrightarrow t$, which follows provided that $F$ itself is
a symmetric
function of its arguments.  This property is in turn guaranteed if the
integrand is symmetric under
$x \leftrightarrow 1-x$, provided, at the same time, we interchange $s$ and
$t$, or
$p_2$ and $p_4$.
The transformation $x\rightarrow 1-x$ can be written as $\theta \rightarrow
\pi/2 - \theta$,
whereupon it is  obvious that we should take
$$
\eqalign{F(s,t) &= \int_0^1 d x x^{-\alpha's-2} (1-x)^{-\alpha' t-2}
\cos(\theta/2-\pi/8)\cr
&= {1\over\sqrt 2}\int_0^1 d x x^{-\alpha's-2} (1-x)^{-\alpha' t-2}\{
\cos\pi/8 \sqrt{1+\sqrt{1+x}} \cr &+ \sin\pi/8 \sqrt{1-\sqrt{1-x}}\}.
\cr}\eqno(4.9)$$

Having found the expression for four tachyon scattering we can examine the
particles
exchanged in a given channel. It is clear from its integral reresentation that
the amplitude obeys duality, that is $F(s,t)$ can be expressed as a sum of
poles in either the s-channel or the t-channel. Let us consider the s-channel.
Expanding the
factor
$(1-x)^{-\alpha't-2}$ we find that
$$
\eqalign{F(s,t)&=\sum_{n=0}^\infty \sum_{p=0}^\infty  \cos{\pi\over8} \sqrt 2
a_p
{(\alpha't+2)(\alpha't +3)\ldots,\alpha't + n + 1)\over(-\alpha's - 1+p+n)}\cr
&+\sum_{n=0}^\infty \sum_{p=0}^\infty  \sin{\pi\over8}  {b_p \over \sqrt 2}
{(\alpha't+2)(\alpha't +3)\ldots,\alpha't + n + 1)\over(-\alpha's - 1+p+n+
1/2)},\cr}
\eqno(4.10)$$
where
$$
\sqrt{1+\sqrt{1+x}} = \sqrt 2 \sum_{p=0}^\infty a_p x^p
\eqno(4.11)$$
and
$$
\sqrt{1-\sqrt{1-x}} = \sqrt{x\over2} \sum_{p=0}^\infty b_p x^p.
 \eqno(4.12)$$
The states exchanged in the first and second terms have masses which satisfy
$\alpha' m^2 = p + n - 1$ and $\alpha' m^2 = p + n - 1/2$ respectively.  While
the former
states are contained in the intercept 1 sector the latter are the
intercept 1/2 sector. Consequently, we find, as discussed above, that
factorisation of the $W_3$ string scattering requires
the existence of intercept 1/2 states in its spectrum in addition to those of
intercept 1 and 15/16 sectors.
The fact that the $W_3$ scattering involves Ising correlations includes the 3
point function. The form of the 3 point functions of the Ising model, as for
any conformal model, are well known to be equivalent to the fusion rules.
Consequently, the fusion rule
$\sigma \sigma = 1 + \epsilon$, where $\sigma$ and $\epsilon$ are the Ising
fields of weight 1/16 and 1/2 respectively implies the existence of intecept
1/2 states in the $W_3$ string.

\par
Following the general tree-level
calculations of $W_3$ string scattering amplitudes
given in [21], covariant approaches to $W_3$ string scattering have been
considered.  The first such attempt [22,26] involved calculations of
correlation functions using only  vertex and picture-changing
operators. The resulting scattering amplitudes disagreed with those of
reference [21] and did not share the connection with Ising model correlation
functions found in [21]; in addition they violated general principles of
S-matrix theory and string theory. The problem was most readily apparent when
considering the scattering of 4 intercept 15/16 strings considered above. It is
easy to see that the vertices with this intercept can never be combined in a
correlator so as to balance the $\varphi$ background charge $2iQ$. As a
consequence, the authors of refences [22,26] concluded that this amplitude and
many like it vanished. The intercept 15/16 amplitude given above, did however
factorise properly [21] and it is clear that this amplitude cannot vanish and
be consistent with factorisation i.e duality [29]. At a more elementary level
it
is clear that the vanishing of such a four point function is inconsistent with
the optical theorem and its generalisations[27,28]
\par
 The use of only vertex and picture-changing
operators to calculate $W_3$ covariant scattering amplitudes was a
straightforward generalization of previous covariant
string scattering calculations for the bosonic and superstrings [9]; in view of
the expected role of $W$-moduli
it is perhaps not surprising that this simple generalization
is not adequate for $W$-strings.

In reference [23], a general covariant procedure for calculating
$W_3$-scattering
was given.  It involved using not only the vertices $V(a,0)$ and the
picture changing operator $P$, but also the screening charge $S$ in the
computation of scattering amplitudes. This method led to
amplitudes in agreement with those found earlier [21] and which satisfied the
general principles expected. We now summarize this approach: the $W_3$ string
scattering amplitudes are to be constructed
from the building blocks
$$
V(a,0)\ ;\ a=1,{15/16},{1/2}\ ;\ S\ ,\ P\eqno(4.13)
$$
and the operation
$$
\int dz\oint\limits_z dv\, b(v).\eqno(4.14)
$$

The latter is a standard operation used for the $b-c$ ghost system.
We must, however, assemble the building blocks so that the $\varphi$
momentum sums to $2iQ$.  This tells us the required number $N_s$ of
screening charges.  We must also have, after carrying out all the
operator product expansions, $3\ c$ ghosts and $5\ e$ ghosts,
otherwise the correlator will vanish.  As usual, we require for an
$N$ string amplitude $N-3$ of the operators of equation (4.14).  The
blocks $V(a,0)$ with $a=1,{15/16}$,  $V({1/2},0)$, $S$ and $P$
have ghosts number $3$, $2$, $-1$ and $1$ respectively.  The ghost number
requirement gives us the number $N_p$ of picture changing operators
$P$.  To be precise if we have a scattering of $N_1$ intercept $1$,
$N_{15/16}$ intercept ${15/16}$ and $N_{1/2}$ intercept
${1/2}$ strings, then $\varphi$ momentum conservation demands
that
$$
8N_1+7N_{15/16}+4N_{1/2}-2N_P = 14\eqno(4.15)
$$
while the ghost number count yields the relations
$$
3N_1+3N_{15/16}+2N_{1/2}-N_S+N_P-
(N_1+N_{15/16}+N_{1/2}-3)=8\eqno(4.16)
$$
These equations imply that
$$
\eqalign{&N_S=4\ N_1+{7\over2}N_{15/16}+2N_{1/2}-7\cr
&N_P=2\ N_1+{3\over2}N_{15/16}+N_{1/2}-2\cr}\eqno(4.17)
$$
\par
To  be concrete let us consider the scattering of 4 intercept 15/16 tachyonic
strings. The above equations tell us that we require 7 factors of $S$ and 4
factors of $P$. One way to distribute these factors in the correlator is as
follows
$$
\eqalign{
\la 0 \vert P & V(15/16,0)(z_1) V( 15/16,1)(z_2)\cr
&  \oint dz_3 \oint_{z_3}dv\,b(v)
  V(15/16,1)(z_3) V(15/16,1)(z_4) S \vert 0 \ra\cr
= & \oint dz_3 \int dw\, \la 0 \vert
  \left( c\,\partial^2 e\,\partial e\, e\ e^{i\beta(15/16,0)\varphi}
  V^x(15/16)\right)(z_1)\cr
& \left( c\,e\ e^{i\beta(15/16,1)\varphi}V^x(15/16)\right)(z_2)
 \left( e\ e^{i\beta(15/16,1)\varphi}V^x(15/16)\right)(z_3)\cr
& \left( c\,e\ e^{i\beta(15/16,1)\varphi}V^x(15/16)\right)(z_4)\cr
&  \left((
d-{5i\over 3\sqrt{58}} \partial b-{2\over
3.87} \partial b\,b\,e-{4i\over 3}{1\over \sqrt{58}} d\,b\,e
)e^{i \beta^s\varphi}\right)(w) \vert 0 \ra\cr
= & - \oint dz_3 \int dw \la 0 \vert
  \left(\prod_{i=1\atop i\ne 3}^4 c(z_i)\right)
  \left(\partial^2 e\,\partial e\,e\right)(z_1)
  e(z_2) e(z_3) e(z_4) d(w)\cr
& e^{i \beta(15/16,0)\varphi}(z_1)
  \prod_{i=2}^4 e^{i \beta(15/16,1)\varphi}(z_i)
  e^{i\beta^s \varphi}(w)
  \prod_{i=1}^4 V^x(15/16)(z_i) \vert 0 \ra\cr
}\eqno(4.18)
$$
By bosonizing the ghosts we find the $c$ ghosts give a factor
$$
(z_1-z_2)(z_1-z_3)(z_2-z_3)\eqno(4.19)
$$
and the $e$-$d$ ghosts a factor
$$
(z_1-w)\prod\limits^4_{i=2}(z_1-z_i)^3(z_i-w)^{-
1}\prod\limits^4_{{i,j=2}\atop {i<j}}(z_i-
z_j).\eqno(4.20)
$$
Evaluating the exponential factors in the usual way we find that the
amplitude for four intercept 15/16 tachyonic states is proportional to
$$
\eqalign{&\int dw\ \int dx\, x(1-x)^{-1/8}[w(1-w)(x-w)]^{-{1\over
4}} (1-x)^{p_2\cdot p_3}x^{p_3\cdot p_4}\cr
=&\int dw\ \int dx\ x^{s/2-2}(1-x)^{t/2-2}[(1-w)(x-w)w]^{-{1\over
4}}\cr}\eqno(4.21)
$$
where we have chosen $z_1=\infty$, $z_2=1$, $z_3=x$ and
$z_4=0$.
This expression can be shown to agree  with that given in equation(4.8), thus
agreeing with the result of reference [21].
\par
Let us now consider the general scattering amplitude and how we must distribute
all the factors of $P$ and $S$, whose number is specified in equation (4.17),
among the vertices so as to gain a non-zero result. To be concrete, let us
consider the scattering of $N_{15 \over16}=2n \geq 6$ intercept 15/16 states,
in which case $N_S=7(n-1)$ and $N_P=3n-2$. One way to do this is to take $n-3$
of the vertices
$V({15/16},2)=(S^2 P)^2V({15/16},0)$ and $n+3$ of the
vertices $V({15/16},1)=S^{2}PV({15/16},0)$.  This leaves
over one $P$ factor and $n-1\ S$ factors and so leads to the
correlator
$$
\eqalign{\la0\vert&\prod\limits^{2n}_{i=4}\left\{\oint
dz_i\oint_{z_i}dv_i\
b(v_i)\right\}\prod\limits^{n+2}_{i=1}V({15/16},1)(z_i)\cr
&P\ V({15/16},1)(z_{n+3})\prod\limits^{2n}_{j=n4}\
V({15/16},2)(z_j)S^{n-1}\vert 0\ra\cr}\eqno(4.22)
$$
We have chosen to assign the final $P$ factor to the $n+3$
vertex, but any other vertex is just as good.  We could also use
two of the final screening charges on the $PV({15/16},1)$
vertex to yield the correlator
$$
\eqalign{&\la 0\vert\prod\limits^{2n}_{i=4}\left\{\oint
dz_i\oint\limits_{z_i}dv_i\
b(v_i)\right\}\prod\limits^{n+2}_{i=1}V({15/16},1)(z_i)\cr
&\prod\limits^{2n}_{j=n+3}V({15/16},2)(z_j)S^{n-
3}\vert 0\ra\cr}\eqno(4.23)
$$
Clearly there are many ways to write such a correlation using also
the vertices $V({15/16},m),m>2$.

For the scattering of $N_{1/2}=2n$ intercept ${1/2}$
states, $N_S=4n-7$ and $N_P=2(n-1)$, we can assign
all the factors of $S$ to the vertices, for example by the choice
of $n-2$ of the vertices $V({1/2},1)$, one of $\bar
V({1/2},0)$ and $n+1$ of the vertices $V({1/2},0)$ and then
place on one of these vertices the required  extra $P$.  The
correlator is then
$$
\eqalign{&\la0\vert\prod\limits^{2n}_{i=4}\left\{\int dz_i\oint dv_i
\,b(v_i)\right\}\prod\limits^{n+1}_{i=1}\ V({1/2},0)(z_i)\cr
&\prod\limits^{2n-1}_{j=n+2} V({1/2},1)(z_j)\ P\bar
V({1/2},0)(z_{2n})\vert 0\ra\cr}\eqno(4.24)$$

Finally for $N_1$ intercept 1 states, $N_S=4N_1-7$ and $N_P=2N_1-
2$, which we can assign as $N_1-2$ vertices $V(1,1)$, one of $V(1,0)$
and one of $\bar V(1,0)$, with one additional factor of $P$ to give
the correlator
$$
\eqalign{\la 0\vert\prod\limits^{N_1}_{i=4}\int
dz_i\oint\limits_{z_i}dv_i\ b(v_i)\prod\limits^{N_1-
2}_{j=1}V(1;1)(z_j)\cr & .P\ V(1,0)(z_{N_1-1})\bar V(1,0)(z_{N_1})\
\vert 0\ra\cr}\eqno(4.25)$$

The many ways of constructing the above correlators lead to the
same results.  The correlator is independent of the place where the
picture changing operator is applied, since
$$
P(z_1)-P(z_2)=[Q,\varphi(z_1)-
\varphi(z_2)]=[Q,\int^{z_2}_{z_1}\partial\varphi],
\eqno(4.26)$$
and this is a BRST trivial operator as it contains
$\partial\varphi$ and not $\varphi$.  Further, we can change on
which vertex the screening charges act by deforming the $w$-contours.
There are also different choices for the contours of the
residual screening charges and these lead to different results.  As
explained above, in the context of the 4 intercept
${15/16}$ scattering, we require these different solutions
since only a particular combination of the contours gives a
crossing symmetric amplitude.

{\bf 5. Discussion}
\par
Let us begin by summarising some of the main results. The physical states of
the $W_3$ string belong to three effective bosonic string sub-sectors of the
theory which have intercepts $a=1$, 1/2 and 15/16 and an effective central
charge of $25+{1 \over 2}$. The count of physical degrees of freedom in the
sub-sector with intercept $a$ involves the Ising model character, $\chi_a$
where $a=1-h$. The cohomology of the BRST charge $Q$ contains an infinite
number of copies of the above states at different ghost numbers and $\varphi$
momenta. There also exists strong evidence that, apart from possible discrete
states, these are the only states in the cohomology of Q.
\par
The scattering of $W_3$ stings, at tree level, has been found to contain, as
part of its integrand, a factor that is none other than the correlation
functions of the Ising model. This result first emerged from an application
[21] of the group theoretic approach to string theory, but it can also be found
using a gauge covariant approach [23].
\par
It would be interesting to give a path-integral derivation of these
scattering results. This would require a
knowledge of $W$-moduli.  Any such derivation would, however, have to
reproduce the above results, and this could provide a clue
to our understanding of $W$-moduli.  We observe that the number of
$W$-moduli is $2 N - 5$ for the scattering of $N$ strings, and this
number emerges from our results in the guise of $N_S - N_P + N_{1/2}$.

\par
The $W_3$ string shows much of the same magic as the bosonic and superstrings;
it obeys a no ghost theorem, it is modular invariant, and the scattering
amplitudes satisfy duality and factorisation. Perhaps the most spectacular
property of the $W_3$ string is its connection with the Ising model. This
connection was first noticed on the basis of phenomological number matching
[7,16], but it also occurs in the count of physical degrees of freedom [10] and
the scattering amplitudes [21,23] and as such it pervade all aspects of
the $W_3$
strings. One can wonder how the $W_3$ string differs from a non-critical string
constructed from the Ising model and D-1 scalar fields $x^{\mu}$. In the latter
theory, the Liouville field will dress the vertices and, as first noticed
in reference [7], can be identified with the remaining scalar. If we adopt the
currently perceived wisdom [30], then the scattering amplitudes for such a
theory will be constructed from such dressed vertices and so will agree with
those found for the $W_3$ string.
\par
Given this correspondence, one is entitled to ask what is the role of the $W$
symmetry when the model can seemingly be formulated without it. Presumably,
although one finds the Ising model in the $W_3$ string one cannot find all
possible tensored Ising models by starting in this way, while for higher $W_N$
strings this statement would extend to tensor products of the unitary minimal
models. One intriguing possibility is that only those  strings constructed from
minimal models that can also be found from $W$, or some other larger symmetry,
are really consistent.
\par
It is interesting to note that the non-perturbative results that emerge from
the matrix model approach to two dimensional quantum gravity do involve $W$
type constraints on the square root of the partition function [38]. Indeed, the
Ising model arises from a two matrix model which involves $W_3$ constraints and
the unitary minimal models which should occur in $W_N$ strings arises from an
$N-1$ matrix model that involves $W_N$ constraints. It would be good to
understand the connections between these results.

\par
The Ising model is usually realised by a Feigin-Fuchs construction [31]
involving only one scalar field, but the $W_3$ string involves an Ising model
constructed from two scalars, the field $\varphi$ and the bosonised $d$, $e$
ghosts. After this talk was given, it was shown [32] that this realisation of
the Ising model involves parafermions, which play a crucial role in determining
the properties of the $W_3$ string.

 {\bf Acknowledgement}
\par
I wish to thank Mike Freeman, with whom many of the results discussed here were
found, as well as Lars Brink, Nathan Berkovits, Bengt Nilsson and Christian
Preitschopf for discussions.

{{\bf References}}

\item{[1]} A. B. Zamolodchikov, Theor. Math. Phys 65 (1989) 1205;
           V.A. Fateev and S. K. Lukyanov, Int. J. Mod. Phys. A3 (1988) 507.
\item{[2]} V. A. Fateev and A. B. Zamolodchikov,
Nucl. Phys. B280 [FS18] (1987) 644.
\item{[3]} L. J. Romans,  Nucl. Phys  B352  (1991) 829.
\item{[4]} M. Kato and K. Ogawa, Nucl. Phys. B212 (1983) 443;
           S. Hwang, Phys. Rev. D10 (1983) 2614.
\item{[5]} A. Neveu, H. Nicolai and P. West; Phys. Lett. 175B (1986) 307
\item{[6]} A. Bilal and J. L. Gervais, Nucl. Phys. B326 (1989) 222.
           P. Howe and P. West, unpublished
\item{[7]} S. Das, A. Dhar and S. Kalyana Rama, Mod. Phys. Lett. A6 (1991);
Mod. Phys. Lett. 269B (1991) 167; Int. J. Mod. Phys. A7 (1992) 2295.
\item{[8]} J. Thierry-Mieg, Phys. Lett. B197 (1987) 368..
\item{[9]} For a review, see D. Lust and S. Theisen, ``Lectures on string
theory,'' Springer-Verlag 1989.
\item{[10]} P. West, Int. J. Mod. Phys. {\bf A8} (1993) 2875.
\item{[11]} E. Del Giudice, P. Di Vecchia and S. Fubini; Ann. Phys. 70 (1972)
378;
	R. C. Brower and P. Goddard; Nucl. Phys. B40 (1972) 437;
	R. C. Brower; Phys. Rev. D6 (1972) 1655;
	P. Goddard and C. Thorn; Phys. Lett. 40B (1972) 235;
        C. B. Thorn; Nucl. Phys. B248 (1984) 551.
\item{[12]}	A. Rocha-Caridi; In Vertex Operations in Mathematics and Physics;
Lepowsky, J et		al (eds), M.S.R.I. publication No.3 p451, Springer-Verlag
(1984).
\item{[13]} A. Capelli, C. Itzykson and J. B. Zuber; Nucl. Phys. B280 [FS18]
(1987) 445.
\item{[14]} E. del Giudice and P. Di Vecchia, Nuovo Cimento 5A (1971) 90.
\item{[15]} M. D. Freeman and D. I. Olive, Phys. Lett. B 175 (1986) 151.
\item{[16]} S. Kalyana Rama, Mod. Phys. Lett {\bf A}6 (1991)3531.
\item{[17]}	C. N. Pope, L. J. Romans, K. S. Stelle; Phys. Lett. 268B (1991)
167, 269B (1991) 287.
\item{[18]}C.N. Pope, S. Schrans and K. W. Xu,
Texas A \& M preprint CTP TAMU-5/92.
\item{[19]} H. Lu, B. E. W. Nilsson, C. N. Pope, K. S. Stelle and P. West; Int.
J. Mod. Phys. {\bf A8} (1993) 4071.
\item{[20]} C. N.  Pope, E. Sezgin, K. S. Stelle and X. J. Wang,
``Discrete states in the $W_3$ string,'' CTP-TAMU-64/92, Imperial/TP/91-92/40.
\item{[21]} M. Freeman and P. West;  Phys. Lett. {\bf B299} (1993) 30.
\item{[22]} H. Lu, C. N. Pope, S. Schrans and X-J. Wang, ``The
interacting $W_3$ string,'' CTP-TAMU-86/92.
\item{[23]} M. Freeman and P. West; Int. J. Mod. Phys. {\bf A8} (1993) 4261.
\item{[24]} For a review see P. West, Nucl. Phys. B (Proc. Suppl) 5B (1988)
217,
Nucl. Phys.
B320 (1989) 103;
 M. Freeman and P. West, Phys. Lett. 205B (1988) 30.
 A. Neveu and P. West, Comm. Math. Phys. 114 (1988) 613.
\item{[25]} A. Luther and I. Peschel, Phys. Rev. B12 (1975) 3908;
A. Belavin, A. Polyakov and A. Zamolodchikov, Nucl. Phys. B241 (1984) 333.
For a review, see C. Itzykson and M. Drouffe, ``Statistical
Field Theory,'' Cambridge University Press.
\item{[26]} H. Lu, C. N. Pope, S. Schrans and X-J. Wang, ``On the spectrum
and scattering of $W_3$ strings,'' CTP-TAMU-4/93.
\item{[27]} See, for example, C. Itzykson and J-B. Zuber,
``Quantum Field Theory,'' McGraw Hill.
\item{[28]} R. Eden, P. Landshoff, D. Olive and J. Polkinghorne,
``The Analytic S-matrix,'' Cambridge University Press.
\item{[29]} R. Dolan, D. Horn and C. Schmid, Phys. Rev. Lett. 19 (1967)
402; Phys. Rev. Lett. 166 (1968) 1768.
\item{[30]} For a review, see N. Ohta, "Discrete States in Two Dimensional
Quantum Gravity", preprint OS-GE-92,  and references therein.
\item{[31]} F. L. Feigin and D. B. Fuchs, Moscow preprint 1983.
Vl. S. Dotsenko and V. A. Fateev, Nucl. Phys. B240 (1984) 312
\item{[32]} M. Freeman and P. West "$W_3$, Strings, Parafermions and the Ising
Model" Kings College and Goteborg preprint; Phys. Lett {\bf B} to be published.
\item{[33]} J. Wess and B. Zumino, Nucl. Phys. B70 (1974)139.
\item{[34]} A. Salam and J. Strathdee, Phys. Rev. D11 (1975) 1521, Nucl. Phys.
B86 (1975)142.
\item{[35]} P. Fayet and J. Illiopoulos Phys. Lett 51B (1974) 461, L.
O'Raifeartaigh, Nucl. Phys. B96 (1975) 331, P. Fayet, Phys. Lett. 58B (1975) 67
\item{[36]} P. West Nucl. Phys. B106 (1976) 219.
\item{[37]} E. Witten, Nucl. Phys. B186 (1981) 150, S. Dimopoulos and H.
Georgi, Nucl. Phys. B193 (1981) 150, N. Sakai, Z. Physik C11 (1982) 153, R.
Kaul Phys. Lett. B109 (1982) 19.
\item{[38]} R. Dijkgraaf, E. Verlinde and H. Verlinde Comm. Math. Phys. 123
(1989) 485, M. Fukuma, H. Kawai and R. Nakayama, Int. J. mod Phys A6 (1991)
1385, J. Goeree, Nucl. Phys. B358 (1991) 737.
\end